\documentclass{svproc}
\usepackage{url}
\usepackage{epsfig}

\begin{document}
\mainmatter              
\title{Optical Properties of Superconducting K$_{0.8}$Fe$_{1.7}$(Se$_{0.73}$S$_{0.27}$)$_2$ Single Crystals}
\titlerunning{Optical Properties of K$_{0.8}$Fe$_{1.7}$(Se$_{0.73}$S$_{0.27}$)$_2$}  
%
\author{Andrei Muratov\inst{1} \and Yevgeny Rakhmanov\inst{1,2} \and
Andrei Shilov\inst{1} \and Igor Morozov\inst{2} \and Yurii Aleshchenko\inst{1}}
\authorrunning{Andrei Muratov et al.} 
%
\tocauthor{Ivar Ekeland, Roger Temam, Jeffrey Dean, David Grove,
Craig Chambers, Kim B. Bruce, and Elisa Bertino}
\institute{P.N. Lebedev Physical Institute of RAS, 119991 Moscow, Russia,\\
\email{aleshchenkoya@lebedev.ru},\\
\and
Lomonosov Moscow State University,
Department of Chemistry, 119991 Moscow, Russia}

\maketitle              

\begin{abstract}
The optical properties of the superconducting K$_{0.8}$Fe$_{1.7}$(Se$_{0.73}$S$_{0.27}$)$_2$ single crystals with a critical temperature $T_c\approx 26$ K have been measured in the {\it ab} plane in a wide frequency range using both infrared Fourier-transform spectroscopy and spectroscopic ellipsometry at temperatures of 4--300 K. The normal-state reflectance of K$_{0.8}$Fe$_{1.7}$(Se$_{0.73}$S$_{0.27}$)$_2$ is analyzed using a Drude-Lorentz model with one Drude component. The temperature dependences of the plasma frequency, optical conductivity, scattering rate, and dc resistivity of the Drude contribution in the normal state are presented. In the superconducting state, we observe a signature of the superconducting gap opening at $2\Delta $(5~K) = 11.8~meV. An abrupt decrease in the low-frequency dielectric permittivity $\varepsilon _1(\omega )$ at $T < T_c$ also evidences the formation of the superconducting condensate. The superconducting plasma frequency $\omega _{pl,s} = (213\pm 5)$~cm$^{-1}$ and the magnetic penetration depth $\lambda =(7.5\pm 0.2)$~$\mu $m at $T=5$~K are determined.
{\sloppy

}
\keywords{iron selenides, high-temperature superconductors, infrared spectroscopy, spectroscopic ellipsometry}
\end{abstract}
\section{Introduction}
After more than fifteen years' extensive studies on iron-based superconductors (IBS), the underlying microscopic mechanisms of both the normal state and superconducting (SC) condensate formation still remain elusive. Similar to high-temperature SC cuprates or heavy fermion systems, the superconductivity of IBS has an unconventional origin characterized by the complex interplay of charge, orbital, and spin degrees of freedom~\cite {Scalapino,Fradkin,Lederer}. Among the IBS, FeSe has the simplest anti-PbO-type crystalline structure. Its critical temperature of SC transition ($T_c$) is about 8~K~\cite {Hsu} but increases up to 37~K under a relatively small hydrostatic pressure of 6~GPa~\cite {Medvedev,Margadonna}. This strong dependence has triggered efforts to find new FeSe-based superconductors by introducing internal chemical pressure~\cite{Wu1}. The search for such materials has resulted in the discovery of a new SC iron-selenide {\it A}$_x$Fe$_2$Se$_2$ family ({\it A} = Na, K, Rb, Cs, K/Tl, Rb/Tl) with $T_c$ as high as 30--35~K~\cite{Guo,Mizuguchi,Krzton-Maziopa2011,Wang1,Ying1,Li,Wang2,Fang,Ying2}. These compounds crystallize in overall simple layered structure with FeSe layers intercalated with alkali metal. They possess a number of unique properties related to the inherent iron deficiency order resulting in the formation of a chiral $\sqrt {5}\times\sqrt {5}\times 1$ superstructure, which reduces the {\it I}4/{\it mmm} crystal symmetry to {\it I}4/{\it m}~\cite{Bacsa}. This iron vacancy ordering is linked with a long-range magnetic order and furthermore, a mesoscopic phase separation is an intrinsic property of the {\it A}$_x$Fe$_2$Se$_2$ family~\cite{Krzton-Maziopa2016}. The majority (80-90\%) phase of the composition {\it A}$_{0.8}$Fe$_{1.6}$Se$_2$ (245) is insulating, magnetic and shows ordered pattern of Fe vacancies in the structure until the lowest available temperatures. The second, minority vacancy-free phase of the composition {\it A}$_x$Fe$_2$Se$_2$ (122) is conducting/semiconducting and becomes superconducting below $T_c\sim 30$--35 K. It has been experimentally proved that tiny domains of the minority phase are uniformly distributed in the AFM-ordered majority 245 phase, forming a specific percolation network throughout the whole macroscopic crystallite, enabling high Meissner screening~\cite {Krzton-Maziopa2021}. The resulting composition {\it A}$_x$Fe$_{2-y}$Se$_2$ is determined as an average over all the coexisting phases. Moreover, angle-resolved photoemission spectroscopy (ARPES) studies revealed that the Fermi surface (FS) topologies of {\it A}$_x$Fe$_{2-y}$Se$_2$ iron selenides are very different from previously known materials. Only the electron pockets are present at the {\it M} corners of the Brillouin zone, while the hole bands at the zone center $\Gamma $ sink below the Fermi level, indicating that the interpocket scattering between the hole and electron pockets is not an essential ingredient for superconductivity~\cite {Zhang,Qian}.

Optical spectroscopy provides valuable information on the bulk electronic properties of superconductors both in the normal and SC state. Shortly after the discovery  of  superconductivity  in  {\it A}$_x$Fe$_{2-y}$Se$_2$ iron selenides, several infrared (IR) studies of these materials have been performed~\cite {Chen,Yuan,Charnukha,Homes1,Ignatov,Wang3,Wu2}. However, the optical data on the SC energy gap and temperature dependences of the normal state parameters of {\it A}$_x$Fe$_{2-y}$Se$_2$ are still scarce with the exclusion of the thorough study performed by Charnukha et al.~\cite {Charnukha}.

In this paper, we report an optical study of K$_{0.8}$Fe$_{1.7}$(Se$_{0.73}$S$_{0.27}$)$_2$ single crystals. To the best of our knowledge, optical studies of this material have not been done earlier.

\section{Experiment}
K$_{0.8}$Fe$_{1.7}$(Se$_{0.73}$S$_{0.27}$)$_2$ single crystals were grown using self-flux technique. The crystals were characterized by x-ray diffraction (XRD), scanning electron microscopy equipped with the energy dispersive x-ray (EDX) spectroscopy, dc resistivity, and magnetotransport measurements. The XRD patterns show sharp (00{\it l}) reflections indicating a high crystalline quality of the samples with the crystallographic {\it c} axis normal to the cleaved surface. The average composition determined by the EDX analysis on 8-10 different positions of the crystal is found to be close to K$_{0.8}$Fe$_{1.7}$(Se$_{0.73}$S$_{0.27}$)$_2$. The samples show a sharp SC transition within 1.3~K with the onset at about 26~K determined as the crossing point of linear $R(T)$ fits of the resistance drop and just above $T_c$, while the d$R(T)/$d$T$ dependence peaks at $T_c\approx 25.6$~K. The details of the sample growth and characterization for the crystals from the same batch of samples can be found elsewhere~\cite {Kuzmicheva1}.

It should be noted that all {\it A}$_x$Fe$_{2-y}$Se$_2$ selenides rapidly degrade in the open air and even in the presence of trace amounts of oxygen or water vapors, with $T_c$ turning to zero in several minutes. For this reason, the samples were transferred from the Ar filled ampoules to the optical cryostats with minimal delay. Freshly cleaved {\it ab}-plane sample surfaces were prepared at ambient conditions using Scotch tape just before pumping the cryostat.

IR measurements in the frequency range of 25--10000~cm$^{-1}$ were performed on a Bruker IFS 125HR Fourier-transform spectrometer equipped with a Si bolometer (IRLabs Inc.) for measurements in the far-IR range and a Konti Spectro A continuous-flow helium cryostat. The sample was mounted on an optically black cone locating at the cold finger of the cryostat. An {\it in situ} aluminum overcoating technique was used to obtain reflectance $R(\omega )$. In order to perform a Kramers-Kronig analysis, we extended the IR reflectance data to the visible and UV range (4000--30 000~cm$^{-1}$) with a Woollam VASE spectral ellipsometer in conjunction with a high-vacuum Janis CRF 725V cryostat. For the low-frequency region, we used a simple $\varepsilon _\infty $ + Drude approximation, since the Hagen-Rubens approximation is applicable only for the samples with a pronounced metallic type of conductivity. All optical measurements were carried out at different temperatures, from 300~K down to 4~K. However, the ellipsometry data prove to be virtually temperature independent.

\begin{figure}[!ht]
\includegraphics[width=8cm]{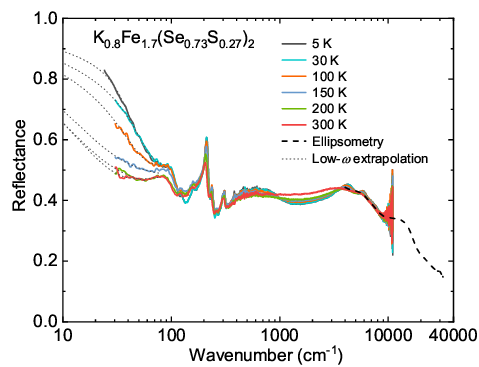}
\centering
\caption{(Color online) The absolute reflectance for K$_{0.8}$Fe$_{1.7}$(Se$_{0.73}$S$_{0.27}$)$_2$ single crystal for light polarized in the {\it ab} plane at several temperatures above and below $T_c$.}
\end{figure}
\section{Results and discussion}
The reflectance spectra of the single crystal of K$_{0.8}$Fe$_{1.7}$(Se$_{0.73}$S$_{0.27}$)$_2$ for a variety of temperatures above and below $T_c$ are shown in Fig.~1. One can see that the overall reflectance is lower than in the case of iron pnictides and roughly close to the value of 0.4. The $R(\omega )$ spectra are dominated by the IR-active phonon vibrations and broad bands at higher energy. For the tetragonal 122 structure having {\it I}4/{\it mmm} crystal symmetry, only two in-plane IR active $E_u$ phonon modes are allowed by the selection rules~\cite {Akrap}. The abundant phonon modes in the {\it ab}-plane IR reflectance of the K$_{0.8}$Fe$_{1.7}$(Se$_{0.73}$S$_{0.27}$)$_2$ single crystals are due to the formation of a $\sqrt {5}\times\sqrt {5}\times 1$ superstructure of Fe vacancies. As a result, the larger unit cell of K$_{0.8}$Fe$_{1.7}$(Se$_{0.73}$S$_{0.27}$)$_2$ has lower symmetry {\it I}4/{\it m} for which already 10 IR-active in-plane modes are allowed~\cite {Bao}. The phonon modes get more narrow and shift to higher frequencies as the temperature is reduced. The low-frequency reflectance starts to increase below 200~K showing a week metallic response. At temperatures approaching $T_c$, the reflectance below 100~cm$^{-1}$ exhibits steep increase.

\begin{figure}[!ht]
\includegraphics[width=8cm]{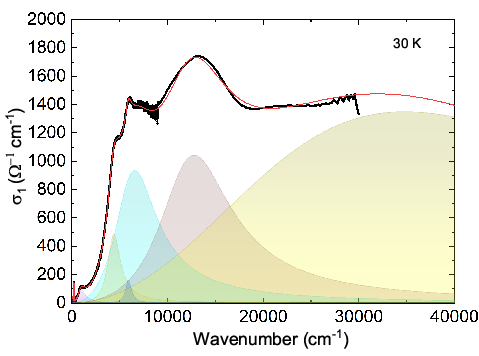}
\centering
\caption{(Color online) The real part of the optical conductivity of K$_{0.8}$Fe$_{1.7}$(Se$_{0.73}$S$_{0.27}$)$_2$ at 30~K over a broad frequency range (dots) together with the Lorentz contributions revealed by the Drude-Lorentz analysis. The red solid curve is a total contour.}
\end{figure}

The optical conductivity $\tilde\sigma $ was obtained by Kramers-Kronig transformation of the $R(\omega)$ spectra. The optical response of K$_{0.8}$Fe$_{1.7}$(Se$_{0.73}$S$_{0.27}$)$_2$ in the normal state was modeled using a Drude-Lorentz model for the complex dielectric function with one Drude component and a set of Lorentz contributions representing the lattice vibrations and interband transitions. According to the Drude-Lorentz model, the complex dielectric function  $\tilde\varepsilon (\omega)=\varepsilon _1(\omega)+i\varepsilon _2(\omega)$ can be represented as
$$
\tilde\varepsilon (\omega)=\varepsilon _{\infty }-\frac{\omega _{D,pl}^2}{\omega (\omega +i\gamma _{D})}+\sum _i\frac{\omega _{i,pl}^2}{\omega _i^2-\omega ^2-i\omega\gamma _i},
$$
where $\varepsilon_\infty $ is the background dielectric function, which comes from contribution of the high frequency absorption bands, $\omega _{D,pl}$ is the Drude plasma frequency, $\gamma _D$ is the scattering rate for the delocalized (Drude) charge carriers, $\omega _{i,pl}$, $\omega _i$ and $\gamma _i$ are the plasma frequency, the center frequency, and the damping of the {\it i}th Lorentz component, respectively. The optical conductivity can be related to the dielectric function as $\tilde\sigma (\omega )=\sigma _1(\omega)+i\sigma _2(\omega )=i\omega [\varepsilon _\infty -\tilde\varepsilon (\omega )]/4\pi $.

As an example, in Fig.~2 we show $\sigma _1(\omega )$ spectra of K$_{0.8}$Fe$_{1.7}$(Se$_{0.73}$S$_{0.27}$)$_2$ at 30~K over a broad frequency range. In addition to the lattice modes (listed in Table~1), there are a number of broad features appeared at high frequency and superimposed on the broad background. The broad bands peaked at $\sim 760$, 4600, 5900, 7100, and 13000~cm$^{-1}$ have been observed earlier in optical studies of the K- and Rb-doped iron selenides~\cite {Chen,Yuan,Charnukha,Homes1} and were ascribed to the spin-controlled interband transitions. We believe that the broad bands at $\sim 1000$, 4400, 5900, 6570, and 12800~cm$^{-1}$ have a similar origin.

\begin{table*}[!ht]
\footnotesize
\centering
\begin{tabular}{|c|c|c|}
  \hline
$\omega _i$ & $\gamma _i$ & $\omega _{i,pl}^2$ \\
  \hline
96.8 & 47.4 & 245.5 \\
  \hline
126.1 & 5.0 & 34.8 \\
  \hline
164.0 & 10.0 & 59.8 \\
  \hline
212.3 & 11.5 & 301.2 \\
  \hline
239.1 & 18.7 & 212.7 \\
  \hline
263.5 & 7.0 & 54.4 \\
  \hline
312.8 & 9.5 & 133.0 \\
  \hline
\end{tabular}
\caption{Physical parameters obtained by fitting the in-plane IR-active lattice modes in the optical conductivity of K$_{0.8}$Fe$_{1.7}$(Se$_{0.73}$S$_{0.27}$)$_2$ at 30~K.}
\end{table*}

\begin{figure}[!ht]
\includegraphics[width=12cm]{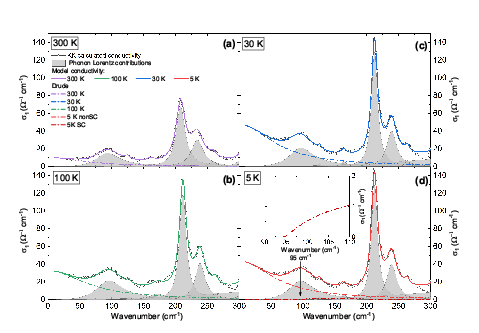}
\centering
\caption{(Color online) The expanded plots of $\sigma _1(\omega )$ spectra below 300~cm$^{-1}$ taken at 300~K (a), 100~K (b), 30~K (c), and 5~K (d) as well as the results of Drude-Lorentz fit to the conductivity of K$_{0.8}$Fe$_{1.7}$(Se$_{0.73}$S$_{0.27}$)$_2$. The Lorentz contributions are shown in gray. The inset in Fig.~3(d) shows a part of the $\sigma _1(\omega )$ data in the region of the SC gapping.}
\end{figure}

Figure 3 shows the optical conductivity spectra of K$_{0.8}$Fe$_{1.7}$(Se$_{0.73}$S$_{0.27}$)$_2$ at different temperatures in the expanded low-frequency region within 300~cm$^{-1}$ as well as the results of a nonlinear least-squares fit to the conductivity using a single Drude component and a set of Lorentz oscillators to reproduce the narrow phonon features and the high-frequency interband transitions. The broad bandwidth of the Lorentz contribution at 96.8~cm$^{-1}$ (see also Table~1) is ascribed to the overlap of at least two IR-active lattice vibrations. There is very small Drude response at low frequency in the conductivity spectrum at 300~K (Fig.~3(a)). With decreasing temperature, the Drude contribution rises. It persists until the lowest temperature (Fig.~3(d)). In this case, it is divided in two contributions. The first, nonSC one, is ascribed to the insulating AFM ordered phase with a $\sqrt {5}\times\sqrt {5}\times 1$ superstructure modulation. The second Drude contribution is related to the SC phase and is gapped at $2\Delta $(5~K) = 95~cm$^{-1}$ (11.8~meV). According to our estimate, the SC phase constitutes not more than 25\% of the sample volume. The temperature dependence of the SC gap is presented in the inset (a) of Fig.~4.

\begin{figure}[!ht]
\includegraphics[width=8cm]{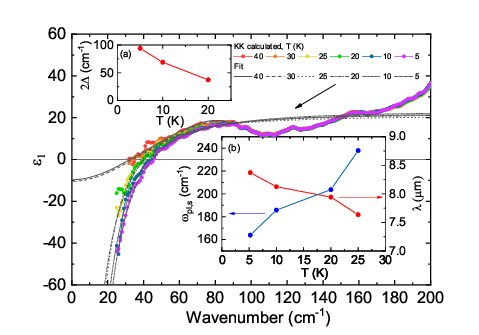}
\centering
\caption{(Color online) The spectra of the dielectric permittivity of K$_{0.8}$Fe$_{1.7}$(Se$_{0.73}$S$_{0.27}$)$_2$ at various temperatures (points) and the fit of the spectra with the expression $\varepsilon _1(\omega )\propto -(\omega _{pl,s}/\omega)^2$ (black lines). The inset (a) shows the temperature dependence of the SC gap while the inset (b) displays the temperature dependences of the SC plasma frequency and the magnetic penetration depth.}
\end{figure}

Another piece of evidence of the SC condensate formation comes from the abrupt decrease in the low-frequency dielectric permittivity $\varepsilon _1(\omega )$ at $T<T_c$. The permittivity spectra of K$_{0.8}$Fe$_{1.7}$(Se$_{0.73}$S$_{0.27}$)$_2$ at various temperatures above and below $T_c$ in Fig.~4 exemplify this behavior. The spectra are fitted with the expression $\varepsilon _1(\omega )\propto -(\omega _{pl,s}/\omega)^2$ (solid line), where $\omega _{pl,s}$ is the SC plasma frequency. This fit allows us to determine $\omega _{pl,s}$. In particular, $\omega _{pl,s}=(213\pm 5)$~cm$^{-1}$ at 5~K. Using the equation for the magnetic penetration depth $\lambda = c/\omega _{pl,s}$~\cite {Basov} ($c$ is the speed of light), we obtain $\lambda =(7.5\pm 0.2)$~$\mu $m at 5~K. This value correlates well with the magnetic penetration depth in K$_{0.8}$Fe$_{2-y}$Se$_2$~\cite {Homes1} and is an order of magnitude higher than what is observed in other iron-arsenide~\cite {Tu} and iron-chalcogenide~\cite {Homes2} superconductors. The temperature dependences of $\omega _{pl,s}$ and $\lambda $ are shown in the inset (b) of Fig.~4. As in the case of the SC gap, we do not extrapolate these dependences to zero temperature to avoid ambiguous values of the parameters related to only a few extrapolation points.

It is interesting to compare the magnitude of the SC gap for K$_{0.8}$Fe$_{1.7}$(Se$_{0.73}$S$_{0.27}$)$_2$ single crystals in the present study (11.8~meV) with that determined in other bulk sensitive experiments with {\it A}$_x$Fe$_{2-y}$Se$_2$ selenides. Optical measurements provide $2\Delta =8$~meV both for Rb$_x$Fe$_{2-y}$Se$_2$~\cite {Charnukha} and K$_{0.8}$Fe$_{2-y}$Se$_2$~\cite {Homes1}. For the K$_{0.8}$Fe$_2$(S$_{0.25}$Se$_{0.75}$)$_2$ single crystals ($T_c=26.9$~K) the value $2\Delta (0)=9.3$~meV was determined directly by intrinsic multiple Andreev reflections effect (IMARE) spectroscopy~\cite {Kuzmicheva2}.
{\sloppy

}

The temperature dependences of the model parameters, such as the plasma frequency $\omega _{pl}$, the static scattering rate $\gamma $, the real part of the optical conductivity $\sigma _1$, and the dc resistivity $\rho $ of the K$_{0.8}$Fe$_{1.7}$(Se$_{0.73}$S$_{0.27}$)$_2$ single crystals in the normal state are shown in Fig.~5. The Drude parameters $\sigma _1=47$~$\Omega ^{-1}\cdot $cm$^{-1}$, $\omega _{pl}=425$~cm$^{-1}$, and $\gamma =64$~cm$^{-1}$ determined from the Drude-Lorentz fit to the optical conductivity at 30~K reveal that the Drude plasma frequency is more than an order of magnitude smaller than that in comparable iron-arsenide~\cite{Tu} or iron-chalcogenide materials~\cite {Homes2}. Notice good agreement between the resistivity data from optical and Van der Pauw measurements.

\begin{figure}[!ht]
\includegraphics[width=8cm]{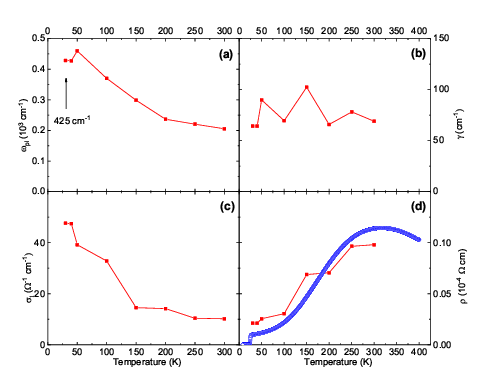}
\centering
\caption{(Color online) The model parameters $\omega _{pl}$, $\gamma $, $\sigma _1$, and $\rho $ of the K$_{0.8}$Fe$_{1.7}$(Se$_{0.73}$S$_{0.27}$)$_2$ single crystals in the normal state as a function of temperature. Blue points in the panel (d) are the results of Van der Pauw measurements.}
\end{figure}

\section{Conclusions}

In conclusion, we have studied the optical properties of the K$_{0.8}$Fe$_{1.7}$(Se$_{0.73}$S$_{0.27}$)$_2$ single crystals in the normal and SC states. A one-Drude-Lorentz model was found to describe successfully the optical properties of K$_{0.8}$Fe$_{1.7}$(Se$_{0.73}$S$_{0.27}$)$_2$ in the normal state. The temperature dependences of the plasma frequency, optical conductivity, scattering rate, and dc resistivity of the Drude contribution in the normal state are presented. There is good agreement between the resistivity data from optical and Van der Pauw measurements. In the SC state, the SC gap $2\Delta $(5~K) = 95~cm$^{-1}$ (11.8~meV) is formed. The volume fraction of the SC phase amounts not more than 25\%. The superconducting plasma frequency $\omega _{pl,s}=(213\pm 5)$~cm$^{-1}$ and the magnetic penetration depth $\lambda =(7.5\pm 0.2)$~$\mu $m are determined at $T=5$~K.
\paragraph{Acknowledgements}
The research was supported by Russian Science Foundation grant No. 22-72-10082-P. The measurements were done using research equipment of the Shared Facilities Center at LPI.
%
%


\begin{thebibliography}{6}
%
\bibitem {Scalapino}
Scalapino, D.J.: A common thread: The pairing interaction for unconventional superconductors. Rev. Mod. Phys. 84, 1383--1417 (2012). \url{doi:10.1103/RevModPhys.84.1383}
%
\bibitem {Fradkin}
Fradkin, E., Kivelson, S.A., and Tranquada, J.M.: Colloquium: Theory of intertwined orders in high temperature superconductors. Rev. Mod. Phys. 87, 457--482 (2015). \url{doi:10.1103/RevModPhys.87.457}
%
\bibitem {Lederer}
Lederer, S., Schattner, Y., Berg, E., and Kivelson, S.A.: Enhancement of Superconductivity near a Nematic Quantum Critical Point. Phys. Rev. Lett. 114, 097001 (2015). \url{doi:10.1103/PhysRevLett.114.097001}
%
\bibitem {Hsu}
Hsu, F.-C., Luo, J.-Y., Yeh, K.-W., Chen, T.-K., Huang, T.-W., Wu, P.M., Lee, Y.-C., Huang, Y.-L., Chu, Y.-Y., Yan, D.-C., and Wu, M.-K.:  Superconductivity in the PbO-type structure $\alpha $-FeSe. Proc. Natl. Acad. Sci. U.S.A. 105, 14262--14264 (2008). \url{doi:10.1073/pnas.0807325105}
%
\bibitem {Medvedev}
 Medvedev, S., McQueen, T.M., Troyan, I.A., Palasyuk, T., Eremets, M.I., Cava, R.J., Naghavi, S., Casper, F., Ksenofontov, V., Wortmann, G., and Felser, C.: Electronic and magnetic phase diagram of $\beta $-Fe$_{1.01}$Se with superconductivity at 36.7~K under pressure. Nat. Mater. 8, 630-633 (2009). \url{doi:10.1038/nmat2491}
%
\bibitem {Margadonna}
Margadonna, S., Takabayashi, Y., Ohishi, Y., Mizuguchi, Y., Takano, Y., Kagayama, T., Nakagawa, T., Takata, M., and Prassides, K.: Pressure evolution of the low-temperature crystal structure and bonding of the superconductor FeSe ($T_c=37$~K). Phys. Rev. B 80, 064506 (2009). \url{doi:10.1103/PhysRevB.80.064506}
%
\bibitem {Wu1}
Wu, M.K., Hsu, F.C., Yeh, K.W., Huang, T.W., Luo, J.Y., Wang, M.J., Chang, H.H., Chen, T.K., Rao, S.M., Mok, B.H., Chen, C.L., Huang, Y.L., Ke, C.T., Wu, P.M., Chang, A.M., Wu, C.T., Perng, T.P.: The development of the superconducting PbO-type $\beta $-FeSe and related compounds. Physica C: Superconductivity 469, 9-12, 340-349 (2009). \url{doi:10.1016/j.physc.2009.03.022}
%
\bibitem {Guo}
Guo, J., Jin, S., Wang, G., Wang, S., Zhu, K., Zhou T., He, M., and Chen, X.: Superconductivity in the iron selenide K$_x$Fe$_2$Se$_2$ ($0\leq x\leq 1.0$), Phys. Rev. B 82, 180520(R) (2010). \url{doi:10.1103/PhysRevB.82.180520}
%
\bibitem {Mizuguchi}
Mizuguchi, Y., Takeya, H., Kawasaki, Y., Ozaki, T., Tsuda, S., Yamaguchi, T., and Takano, Y.: Transport properties of the new Fe-based superconductor K$_x$Fe$_2$Se$_2$ ($T_c = 33$~K). Appl. Phys. Lett. 98, 042511 (2011). \url{doi:10.1063/1.3549702}
%
\bibitem {Krzton-Maziopa2011}
Krzton-Maziopa, A., Shermadini Z., Pomjakushina, E., Pomjakushin, V., Bendele, M., Amato, A., Khasanov, R., Luetkens, H., and Conder, K.: Synthesis and crystal growth of Cs$_{0.8}$(FeSe$_{0.98}$)$_2$: a new iron-based superconductor with $T_c = 27$~K. J. Phys.: Condens. Matter 23, 052203 (2011). \url{doi:10.1088/0953-8984/23/5/052203}
%
\bibitem{Wang1}
Wang, A.F., Ying, J.J., Yan, Y.J., Liu, R.H., Luo, X.G., Li, Z.Y., Wang, X.F., Zhang, M., Ye, G.J., Cheng, P., Xiang, Z.J., and Chen, X.H.: Superconductivity at 32~K in single-crystalline Rb$_x$Fe$_{2-y}$Se$_2$. Phys. Rev. B 83, 060512(R) (2011). \url{doi:10.1103/PhysRevB.83.060512}
%
\bibitem {Ying1}
Ying, J.J., Wang, X.F., Luo, X.G., Wang, A.F., Zhang, M., Yan, Y.J., Xiang, Z.J., Liu, R.H., Cheng, P., Ye, G.J., and Chen, X.H.:  Superconductivity and magnetic properties of single crystals K$_{0.75}$Fe$_{1.66}$Se$_2$ and Cs$_{0.81}$Fe$_{1.61}$Se$_2$. Phys. Rev. B 83, 212502 (2011). \url{doi:10.1103/PhysRevB.83.212502}
%
\bibitem {Li}
Li, C.-H., Shen, B., Han, F., Zhu, X., and Wen, H.-H.: Transport properties and anisotropy of Rb$_{1-x}$Fe$_{2-y}$Se$_2$ single crystals. Phys. Rev. B 83, 184521 (2011). \url{doi:10.1103/PhysRevB.83.184521}
%
\bibitem {Wang2}
Wang, H.-D., Dong, C.-H., Li, Z.-J., Mao, Q.-H., Zhu, S.-S., Feng, C.-M., Yuan, H.Q., and Fang, M.-H.: Superconductivity at 32~K and anisotropy in Tl$_{0.58}$Rb$_{0.42}$Fe$_{1.72}$Se$_2$ crystals. Europhys. Lett. 93, 47004 (2011). \url{doi:10.1209/0295-5075/93/47004}
%
\bibitem {Fang}
Fang, M.-H., Wang, H.-D., Dong, C.-H., Li, Z.-J., Feng, C.-M., Chen, J., and Yuan, H.Q.: Fe-based superconductivity with $T_c=31$~K bordering an antiferromagnetic insulator in (Tl,K)Fe$_x$Se$_2$. Europhys. Lett. 94, 27009 (2011). \url{doi:10.1209/0295-5075/94/27009}
%
\bibitem {Ying2}
Ying, T.P., Chen, X.L., Wang, G., Jin, S.F., Zhou, T.T., Lai, X.F., Zhang, H., and Wang, W.Y.: Observation of superconductivity at 30$\sim $46K in A$_x$Fe$_2$Se$_2$ (A = Li, Na, Ba, Sr, Ca, Yb, and Eu). Sci. Rep. 2, 426 (2012). \url{doi:10.1038/srep00426}
%
\bibitem {Bacsa}
Bacsa, J., Ganin, A.Y., Takabayashi, Y., Christensen, K.E., Prassides, K., Rosseinsky, M.J., and Claridge, J.B.: Cation vacancy order in the K$_{0.8+x}$Fe$_{1.6-y}$Se$_2$ system: Five-fold cell expansion accommodates 20\% tetrahedral vacancies. Chem. Sci. 2, 1054--1058 (2011). \url{doi:10.1039/C1SC00070E}
%
\bibitem {Krzton-Maziopa2016}
Krzton-Maziopa, A., Svitlyk, V., Pomjakushina, E., Puzniak, R., and Conder, K.: Topical review: Superconductivity in alkali metal intercalated iron selenides. J. Phys.: Condens. Matter 28, 293002 (2016). \url{doi:10.1088/0953-8984/28/29/293002}
%
\bibitem {Krzton-Maziopa2021}
Krzton-Maziopa, A.: Intercalated Iron Chalcogenides: Phase Separation Phenomena and Superconducting Properties. Front. Chem. 9, 640361 (2021). \url{doi:10.3389/fchem.2021.640361}
%
\bibitem {Zhang}
Zhang, Y., Yang, L.X., Xu, M., Ye, Z.R., Chen, F., He, C., Jiang, J., Xie, B.P., Ying, J.J., Wang, X.F., Chen, X.H., Hu, J.P., and Feng, D.L.: Nodeless superconducting gap in A$_x$Fe$_2$Se$_2$ (A=K,Cs) revealed by angle-resolved photoemission spectroscopy. Nat. Mater. 10, 273--277 (2011). \url{doi:10.1038/nmat2981}
%
\bibitem {Qian}
Qian, T., Wang, X.-P., Jin, W.-C., Zhang, P., Richard, P., Xu, G., Dai, X., Fang, Z., Guo, J.-G., Chen, X.-L., and Ding, H.: Absence of a Holelike Fermi Surface for the Iron-Based K$_{0.8}$Fe$_{1.7}$Se$_2$ Superconductor Revealed by Angle-Resolved Photoemission Spectroscopy. Phys. Rev. Lett. 106, 187001 (2011). \url{doi:10.1103/PhysRevLett.106.187001}
%
\bibitem {Chen}
Chen, Z.G., Yuan, R.H., Dong, T., Xu, G., Shi, Y.G., Zheng, P., Luo, J.L., Guo, J.G., Chen, X.L., and Wang, N.L.: Infrared spectrum and its implications for the electronic structure of the semiconducting iron selenide K$_{0.83}$Fe$_{1.53}$Se$_2$. Phys. Rev. B 83, 220507(R) (2011). \url{doi:10.1103/PhysRevB.83.220507}
%
\bibitem {Yuan}
Yuan, R.H., Dong, T., Song, Y.J., Zheng, P., Chen, G.F., Hu, J.P., Li, J.Q., and Wang, N.L.: Nanoscale phase separation of antiferromagnetic order and superconductivity in K$_{0.75}$Fe$_{1.75}$Se$_2$. Scientific Reports 2, 221 (2012). \url{doi:10.1038/srep00221}
%
\bibitem {Charnukha}
Charnukha, A., Deisenhofer, J., Pr\"opper, D., Schmidt, M., Wang, Z., Goncharov, Y., Yaresko, A.N., Tsurkan, V., Keimer, B., Loidl, A., and Boris, A.V.: Optical conductivity of superconducting Rb$_2$Fe$_4$Se$_5$ single crystals. Phys. Rev. B 85, 100504(R) (2012). \url{doi:10.1103/PhysRevB.85.100504}
%
\bibitem {Homes1}
Homes, C.C., Xu, Z.J., Wen, J.S., and Gu, G.D.: Optical conductivity of superconducting K$_{0.8}$Fe$_{2-y}$Se$_2$ single crystals: Evidence for a Josephson-coupled phase. Phys. Rev. B 85, 180510(R) (2012). \url{doi:10.1103/PhysRevB.85.180510}
%
\bibitem {Ignatov}
Ignatov, A., Kumar, A., Lubik, P., Yuan, R.H., Guo, W.T., Wang, N.L., Rabe, K., and Blumberg, G.: Structural phase transition below 250~K in superconducting K$_{0.75}$Fe$_{1.75}$Se$_2$. Phys. Rev. B 86, 134107 (2012). \url{doi:10.1103/PhysRevB.86.134107}
%
\bibitem {Wang3}
Wang, Z., Schmidt, M., Fischer, J., Tsurkan, V., Greger, M., Vollhardt, D., Loidl, A., and Deisenhofer, J.: Orbital-selective metal-insulator transition and gap formation above $T_c$ in superconducting Rb$_{1-x}$Fe$_{2-y}$Se$_2$. Nat. Commun. 5, 3202 (2014). \url{doi:10.1038/ncomms4202}
%
\bibitem {Wu2}
Wu, M.K., Wu, P.M., Wen, Y.C., Wang, M.J., Lin, P.H., Lee, W.C., Chen, T.K., and Chang, C.C.: An overview of the Fe-chalcogenide Superconductors, J. Phys. D: Appl. Phys. 48, 323001 (2015). \url{doi:10.1088/0022-3727/48/32/323001}
%
\bibitem {Kuzmicheva1}
Kuzmicheva, T., Morgun, L., Gavrilkin, S., Kuzmichev, S., Degtyarenko, A., Muratov, A., Zhuvagin, I., Shilov, A., Rakhmanov Y., and Morozov, I.: Single Crystal Growth, Transport Phenomena, and Upper Critical Field of Alkali Metal-Based K$_x$Fe$_{2-y}$(Se,S)$_2$ Iron Chalcogenides. J. Supercond. Nov. Magn. 38, 120 (2025). \url{doi:10.1007/s10948-025-06963-2}
%
\bibitem {Akrap}
Akrap, A., Tu, J.J., Li, L.J., Cao, G.H., Xu, Z.A., and Homes, C.C.: Infrared phonon anomaly in BaFe$_2$As$_2$, Phys. Rev. B 80, 180502(R) (2009). \url{doi:10.1103/PhysRevB.80.180502}
%
\bibitem {Bao}
Bao, W., Huang, Q.-Z., Chen, G.-F., Green, M.A., Wang, D.-M., He, J.-B., and Qiu, Y.-M.: A Novel Large Moment Antiferromagnetic Order in K$_{0.8}$Fe$_{1.6}$Se$_2$ Superconductor. Chin. Phys. Lett. 28, 086104 (2011). \url{doi:10.1088/0256-307X/28/8/086104}
%
\bibitem {Basov}
Basov, D.N., Timusk, T.: Electrodynamics of high-$T_c$ superconductors. Rev. Modern Phys. 77, 721--779 (2005). \url{doi:10.1103/RevModPhys.77.721}
%
\bibitem {Tu}
Tu, J.J., Li, J., Liu, W., Punnoose, A., Gong, Y., Ren, Y.H., Li, L.J., Cao, G.H., Xu, Z.A., and Homes, C.C.: Optical properties of the iron arsenic superconductor BaFe$_{1.85}$Co$_{0.15}$As$_2$. Phys. Rev. B 82, 174509 (2010). \url{doi:10.1103/PhysRevB.82.174509}
%
\bibitem {Homes2}
Homes, C.C., Akrap, A., Wen, J.S., Xu, Z.J., Lin, Z.W., Li, Q., and Gu, G.D.: Electronic correlations and unusual superconducting response in the optical properties of the iron chalcogenide FeTe$_{0.55}$Se$_{0.45}$. Phys. Rev. B 81, 180508(R) (2010). \url{doi:10.1103/PhysRevB.81.180508}
%
\bibitem {Kuzmicheva2}
Kuzmicheva, T.E., Kuzmichev, S.A., Ilina, A.D., Nikitchenkov, I.A., Rakhmanov, Ye.O., Shilov, A.I., and Morozov, I.V.: Comparison of the Superconducting Properties of Potassium Iron Selenides with Isovalent Substitution, JETP Letters, Vol. 121, No. 8, pp. 662-669 (2025). \url{doi:0.1134/S0021364025605792}
%

\end{thebibliography}
\end{document}